\documentclass[preprint,12pt]{elsarticle}

\usepackage{amssymb}
\usepackage{amsmath}

\usepackage{float}
\usepackage{multirow}
\usepackage{xr}
\usepackage{hyperref}

\usepackage{subcaption}
\usepackage{graphicx}

\usepackage{xcolor}

\journal{Energy and AI}

\begin{document}

\begin{frontmatter}



\title{A Hybrid Artificial Intelligence Method for Estimating Flicker in Power Systems} 

\author[a]{Javad Enayati}
\author[b]{Pedram Asef}
\author[d]{Alexandre Benoit} 

\affiliation[a]{organization={R\&D Department, Sander Elektronik AG},
            addressline={Stauseestrasse 73}, 
            city={Böttstein},
            postcode={5314},
            state={Zurich},
            country={Switzerland}}

\affiliation[b]{organization={Department of Mechanical Engineering, University College London (UCL)},
            addressline={ Torrington Place}, 
            city={London},
            postcode={WC1E 7JE},
            country={UK}}

\affiliation[d]{organization={Engineering Department, University of Cambridge},
            addressline={Trumpington St},
            city={Cambridge},
            postcode={CB2 1PZ}, 
            state={Cambridgeshire},
            country={UK}}

\begin{abstract}

This paper introduces a novel hybrid method combining H-$\infty$ filtering 
and an adaptive linear neuron (ADALINE) network for flicker component 
estimation in power distribution systems. The proposed method leverages 
the robustness of the H-$\infty$ filter to extract the voltage envelope 
under uncertain and noisy conditions, followed by the use of ADALINE to 
accurately identify the relative amplitudes of flicker components 
($\Delta V_i / V_t$) at standard IEC-defined frequencies embedded in the 
envelope. This synergy enables efficient time-domain estimation with rapid 
convergence and noise resilience, addressing key limitations of existing 
frequency-domain approaches. Unlike conventional techniques, this hybrid 
model handles complex power disturbances without prior knowledge of noise 
characteristics or extensive training. To validate the method's performance, 
we conduct simulation studies based on IEC Standard 61000-4-15, supported 
by statistical analysis, Monte Carlo simulations, and real-world data. 
Results demonstrate superior accuracy, robustness, and reduced computational 
load compared to Fast Fourier Transform (FFT) and Discrete Wavelet Transform 
(DWT)-based estimators.
\end{abstract}

\begin{graphicalabstract}
\begin{figure}
    \centering
    \includegraphics[width=1\linewidth]{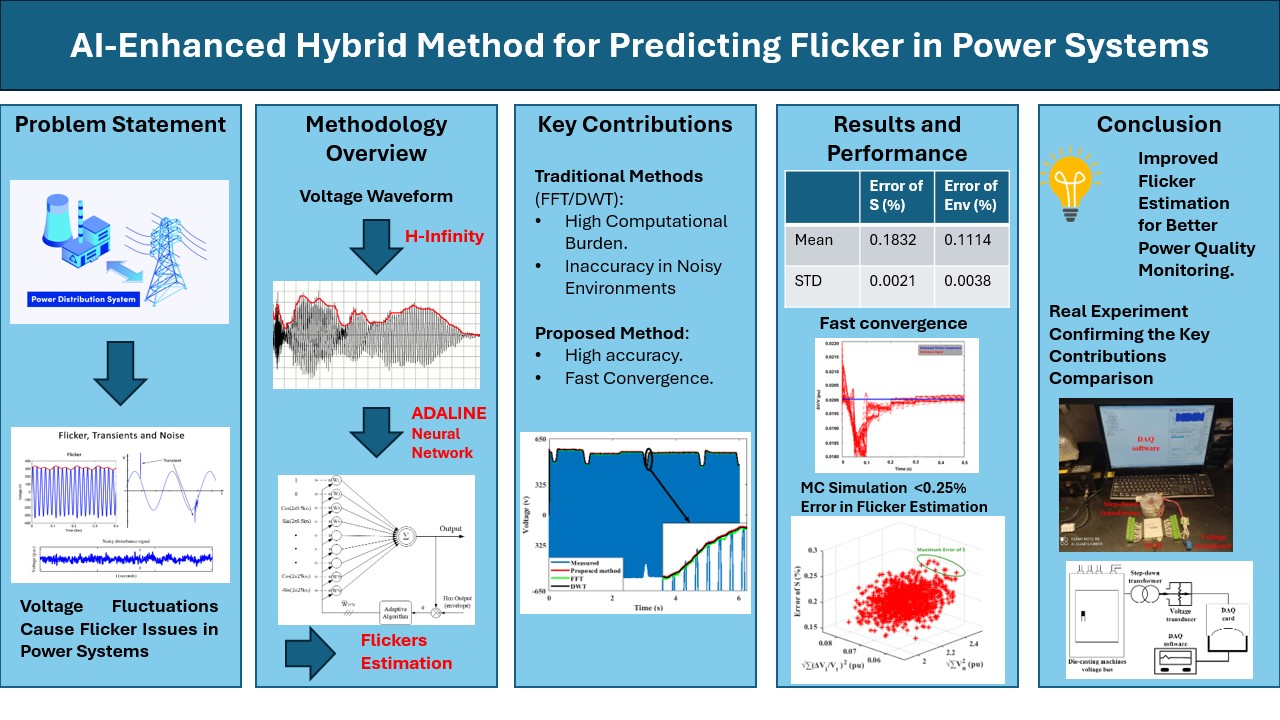}    
    \label{figGA}
\end{figure}
\end{graphicalabstract}

\begin{highlights}
\item Development of AI-driven hybrid method using H-\(\infty\) and Artificial Neural Networks techniques for accurate flicker prediction.
\item Introduction of a novel voltage fluctuation model that incorporates both harmonic and flicker components, offering a more accurate representation of power signals.
\item The proposed hybrid algorithm separates carrier and flickers frequencies with differing amplitudes, achieving fast convergence without traditional filtering requirements.
\end{highlights}


\end{frontmatter}



\section{Introduction}
\label{sec1}

Voltage fluctuation refers to a series of changes or continuous variations in the root mean square (RMS) or peak value of the voltage. According to power quality standards, the magnitude of these fluctuations typically remains within 10\% of the nominal voltage \cite{b1}\cite{b2}. Voltage fluctuations can cause light flicker, which is perceived as unsteady luminance or variations in the spectral distribution of light sources. These issues are particularly relevant in renewable energy systems, where the variability of wind and solar generation often introduces dynamic and unpredictable voltage changes that contribute to flicker phenomena \cite{b2_bis, b2_bisbis}. It should be noted that while voltage fluctuations are the primary source 
of flicker phenomena in power systems, the perceptibility of flicker also 
depends strongly on the characteristics of the lighting technology. 
Incandescent lamps, due to their thermal inertia, attenuate high-frequency 
voltage fluctuations, whereas fluorescent and LED lamps—particularly those 
with low-quality driver circuits—may translate even small disturbances into 
visible luminance variations \cite{b6, b41}. 
Furthermore, built-in electronic buffers or power factor correction 
mechanisms in modern lamps can mitigate or exacerbate flicker sensitivity. 
Thus, voltage fluctuation is a necessary but not sufficient condition for 
light flicker; the overall effect results from the interaction of the 
electrical disturbance with lamp dynamics and human visual response, as 
formalized in the IEC 61000-4-15 standard.
To assess this flicker, the IEC flicker meter incorporates the human eye-brain response to characterize light variations. The key output of the IEC flicker meter is the instantaneous flicker sensation (S). The IEC 61000-4-15 standard defines four analytical blocks to calculate flicker sensation \cite{b3, b4, b5, b6}. The first block normalizes the input voltage waveform relative to an internal reference level. In the second block, the output is squared. Block 3 then applies demodulation filters, including a high-pass filter (cut-off at 0.05 Hz) and a low-pass filter (cut-off at 35 Hz), along with a weighting filter. Finally, Block 4 computes the S parameter by squaring the signal and using a divider to normalize the results based on the mean value.

While the IEC approach involves continuous voltage input for flicker calculation, the filtering process introduces a high computational burden in real-world applications. As a result, several alternative methods using discretized waveforms have been proposed. Many of these rely on the discrete Fourier transform (DFT), a frequency-domain approach \cite{b7}\cite{b8}\cite{b9}. However, DFT-based methods face challenges such as aliasing, picket fence effects, and leakage phenomena, all of which can significantly reduce accuracy in the presence of noise and disturbances \cite{b10}. To mitigate these issues, some researchers have proposed frequency-domain approaches using the Z-transform to compensate for leakage errors \cite{b11}\cite{b12}. In addition to Z-transform approaches, advanced techniques like Dynamic Voltage Restorers (DVR) with Multi-Level Inverters (MLI) have been used to mitigate voltage sags, swells, and flickers. Novel configurations, such as a 33-level Asymmetrical MLI, enhance Low and High Voltage Ride Through (LVRT, HVRT) capabilities, showing improved performance in compensating for flicker and harmonic issues \cite{newref1}.
 Despite acceptable performance in certain conditions, these methods still struggle with poor accuracy in presence of noise and harmonics.

To address these limitations, a combination of Z-transform and the Teager energy operator has been suggested \cite{b13, b14}, offering improved accuracy but only at high sampling rates, which increases computational complexity. Moreover, these methods often consider only a limited number of harmonics and flicker components in simulations. Given the shortcomings of frequency-domain processors, extensive research has been conducted on time-domain processing techniques \cite{b15}\cite{b16}, which are generally used to estimate harmonic components in power systems \cite{b17, b18, b19, b20}. Since flicker components manifest as frequency components near the fundamental frequency, the applied methods must be robust against close-frequency interactions to accurately detect flicker components.

In \cite{b21, b22, b23, b24}, amplitude modulation (AM) is studied for its ability to estimate flicker components in distributed generation (DG) systems using multiple filters. While this approach offers simplicity, its accuracy heavily depends on the number of filters used, and practical applications may be hindered by high computational time when detailed flicker waveforms are required. Analytical evaluations of flicker meter performance based on IEC 61000-4-15 are documented in \cite{b25, b26, b27, b28}. These studies use RMS voltage inputs in offline mode, which is suitable for planning but limited in real-time applications. Although the results are validated through both simulations and field tests, flicker meters tend to be accurate for low-frequency inputs but unreliable for higher-frequency fluctuations \cite{b26}.

Artificial neural networks (ANNs) are powerful tools for optimization and estimation tasks and have been applied to flicker detection in power systems \cite{b29, b30}. However, the number of network inputs tends to increase when ANNs are directly used for flicker index extraction, leading to higher computational complexity. To mitigate this, multi-stage networks with pre-processing stages have been introduced \cite{b31}. Although these approaches yield acceptable results in certain scenarios, they are mostly limited to offline applications.

The Kalman filter (KF) and its variants have been widely used as fundamental tools for analyzing and solving various estimation problems \cite{b22, b23}. Several studies have utilized state variable representations of nonlinear flicker waveforms to extract flicker components using different types of KFs \cite{b34, b35, b36, b37}. While KF-based methods provide reliable outputs when targeting a limited number of flicker frequencies, their accuracy diminishes as the frequency range broadens. Additionally, simple KF algorithms are prone to instability when noise and fluctuations are present in the reference signal.

In this paper, we propose a novel time-domain hybrid method called the Hybrid Envelope Flicker System (HEFS) to estimate flicker components from modulated voltage waveforms in power systems \cite{ourpaper}. The algorithm operates in two stages: first, the envelope of the waveform is extracted using a robust H-$\infty$ filter \cite{newref2}; second, an online neural network (such as ADALINE) is applied to estimate the flicker components \cite{newref3, newref4, newref5}. This method benefits from a simple formulation and robust performance. The main contributions of this work are summarized as follows:

Despite extensive research on flicker estimation, most methods struggle to 
adapt reliably under non-sinusoidal disturbances where harmonics, noise, and 
complex modulation distort the voltage waveform. Existing ANN-based models 
often demand high input dimensionality and extensive training datasets, 
while Kalman filters degrade under model uncertainty or require restrictive 
assumptions. These limitations motivate the development of a method that is 
both robust to uncertainty and computationally efficient for real-time 
applications. In this work, we specifically address disturbances arising from voltage flicker, harmonic distortion, and additive white noise, which frequently coexist in real-world power systems. The main contributions of this paper are summarized below:

\begin{enumerate}
    \item Employs a robust H-$\infty$ estimator with a simple formulation, capable of detecting flicker in modulated voltage waveforms, even in the presence of unknown noise disturbances.
    \item Utilizes the ADALINE neural network \cite{b38}, a fast estimator capable of handling multiple envelopes without requiring a training phase, providing an efficient method for estimating voltage fluctuations in the presence of power harmonics.
    \item Introduces a novel voltage fluctuation model that incorporates both harmonic and flicker components, offering a more accurate representation of power signals.
    \item Extracts multiple flicker frequencies from a fluctuating voltage waveform, capturing all critical frequency components as defined by the IEC 61000-4-15 standard.
    \item Proposes a hybrid algorithm that separates carrier and flicker frequencies with differing amplitudes, achieving fast convergence without the need for traditional filtering methods.
\end{enumerate}

In summary, the proposed hybridization of H-$\infty$ filtering with 
ADALINE provides a novel two-stage approach that enables accurate, 
real-time estimation of both harmonic and flicker components in compliance 
with IEC 61000-4-15.




\section{Methods}

In the IEC standard, voltage waveforms with flicker components are represented by applying an AM signal, where the power system frequency (50 or 60 Hz) serves as the carrier frequency, and the flicker frequency (envelope) acts as the message frequency. This relationship is expressed in discrete form \cite{b2}:

\begin{equation}
Z^k = V_c \cos(2\pi f k \tau_s) \left\{ 1 + \sum_{i=1}^{F} \frac{\Delta V_i}{2V} \cos(2\pi F_i k \tau_s + \theta_i) \right\}
\label{eq1}
\end{equation}

where \(V_c\) is the carrier amplitude (nominal voltage amplitude), \( f \) is the power system frequency, and the index \( i = 1, 2, \dots, F \) refers to each flicker component. \( F_i \) and \( \theta_i \) represent the frequency and phase angle of the \( i \)-th flicker component, respectively, while \( \frac{\Delta V_i}{2V} \) denotes the relative voltage fluctuation. The sampling period \( \tau_s \) is used to implement the model in discrete form. The goal of the proposed algorithm is to estimate the relative voltage amplitude \( \frac{\Delta V_i}{2V} \) for each flicker component. It should be noted that in all equations, the superscript \( k \) represents the time point in discrete form.

Due to the presence of non-linear loads, the voltage waveform in power systems becomes distorted. The previously used model Equation~\ref{eq1} does not fully capture the complexities of real power signals. Therefore, this paper introduces a novel voltage fluctuation model that accounts for both harmonic and flicker components. Additionally, to model the noise properties more accurately, a high-frequency noise component is incorporated into the signal model as follows:

\begin{equation}
Z^k = \sum_{n=1}^{N} V_n \cos(2\pi n f k \tau_s + \phi_n) \left\{ 1 + \sum_{i=1}^{F} \frac{\Delta V_i}{2V_t} \cos(2\pi F_i k \tau_s + \theta_i) \right\} + \sigma \cdot \text{randn}_k
\label{eq2}
\end{equation}

where \( n = 1, 2, \dots, N \) denotes the order of harmonics. \( V_n \) represents the amplitude of each harmonic, and \( \phi_n \) is the corresponding phase. \( \sigma \cdot \text{randn}_k \) represents Gaussian noise, where the standard deviation \( \sigma \) defines the noise power. The value \( V_t = \sqrt{\sum_{n=1}^{N} V_n^2} \) serves as the base value for normalizing the voltage fluctuations.

Given the large number of states with varying amplitude magnitudes in the model, improved performance in flicker estimation is achieved through the use of hybrid algorithms. The proposed hybrid algorithm operates in two sequential stages. First, the envelope of the fluctuating voltage waveform is extracted using the H-$\infty$ filter at each time step \( k \). The extracted envelope is then passed to the ADALINE network, which estimates the \( \frac{\Delta V_i}{V_t} \) values corresponding to each frequency at that time.

\subsection{Envelope Estimation Using H-$\infty$ Method}

The H-$\infty$ filter is a widely-used analytical estimator that minimizes the worst-case estimation error \cite{b32}. It performs efficiently in the estimation of time-varying signals, even in the presence of unknown noise, contrasting with KF-based methods, which minimize the expected variance of the estimation error. To estimate the envelopes using the H-$\infty$ filter, the modulated signal in Equation~\ref{eq2} is reformulated as follows:

\begin{equation}
Z^k = A_1\{ \text{Envelope}_1 \} + A_2\{ \text{Envelope}_2 \} + \dots + A_N\{ \text{Envelope}_N \} + \sigma \cdot \text{randn}^k
\label{eq3}
\end{equation}

where:

\begin{equation}
\text{Envelope}_n = V_n + \sum_{i=1}^{F} \frac{V_n \Delta V_i}{2V_t} \cos(2\pi F_i k \tau_s + \theta_i)
\label{eq4}
\end{equation}

The power frequency components, i.e., \( A_n = \cos(2\pi n f k \tau_s + \phi_n) \), are expanded using trigonometric identities:

\begin{equation}
A_n = \cos(2\pi n f k \tau_s) \cos(\phi_n) - \sin(2\pi n f k \tau_s) \sin(\phi_n)
\label{eq5}
\end{equation}

To capture the true power frequency, particularly in cases of frequency deviation issues, a Phase-Locked Loop (PLL) from the MATLAB toolbox is applied. Based on Equation~\ref{eq3}, a linear model for the H-$\infty$ estimator is derived:

\begin{equation}
Z^k = H^k x + \omega^k
\label{eq6}
\end{equation}

where \( \omega_k \) is the measurement noise matrix, and the state vector \( x \) is defined as:

\begin{equation}
\begin{split}
    x = [\cos(\phi_1) \text{Envelope}_1 \sin(\phi_1) \text{Envelope}_1 \dots \\ 
    \cos(\phi_N) \text{Envelope}_N \sin(\phi_N) \text{Envelope}_N]
\end{split}
\label{eq7}
\end{equation}

The state vector \( x \) contains information of the envelopes. The structure matrix \( H^k \) is:

\begin{equation}
H^k = [\cos(2\pi f k \tau_s) - \sin(2\pi f k \tau_s) \dots \cos(2\pi N f k \tau_s) - \sin(2\pi N f k \tau_s)]
\label{eq8}
\end{equation}

The state vector \( x \) is estimated in the time domain using the measurement \( Z^k \), following the time-update equation:

\begin{equation}
\hat{x}^k = \Phi (\hat{x}^{k-1} + K^k (Z^k - H^k \hat{x}^{k-1}))
\label{eq9}
\end{equation}

where \( K^k \) and \( \Phi \) represent the gain and state transition matrices, respectively. Since the time-varying characteristics of the states are unknown, \( \Phi \) is assumed to be a constant identity matrix. The covariance matrix of the estimation error is updated recursively by minimizing the worst-case estimation error:

\begin{equation}
P^k = \Phi P^{k-1}(I - \alpha P^{k-1} + {H^k}^T R^{-1} H^k P^{k-1})^{-1} \Phi^T
\label{eq10}
\end{equation}

where the factor \( \alpha \) controls the influence of previous errors in the estimation process and is chosen by the filter designer based on the specific problem. \( R \) represents the covariance matrix of the measurement noise. The gain matrix \( K^k \) is computed using the covariance matrix of the estimation error:

\begin{equation}
K^k = P^k (I - \alpha P^k + {H^k}^T R^{-1} H^k P^k)^{-1} {H^k}^T R^{-1}
\label{eq11}
\end{equation}

At each time step \( k \), the envelopes can be obtained using information from the states:

\begin{equation}
\text{Envelope}_1 = \sqrt{x_1^2 + x_2^2}, \dots, \text{Envelope}_n = \sqrt{x_{N-1}^2 + x_N^2}
\label{eq12}
\end{equation}

To summarize, the H-$\infty$ is a robust state estimator designed to minimize the worst-case estimation error over all possible disturbances, rather than minimizing the expected error covariance as in the Kalman filter. Its optimization objective is to guarantee that the ratio between the estimation error energy and the disturbance energy remains below a prescribed threshold $\gamma$, i.e.,

\begin{equation}
    \frac{\| e_k \|_2^2}{\| \omega_k \|_2^2} < \gamma^2
\end{equation}

where $e_k$ is the estimation error and $\omega_k$ is the measurement noise. This formulation ensures robustness even when the exact noise statistics are unknown. In practice, the filter recursively updates the state estimate by incorporating measurement innovations while bounding the amplification of disturbances. The optimization problem is solved by computing a gain matrix $K_k$ (see above) which ensures convergence under uncertain and noisy environments. This makes the H-$\infty$ framework especially suitable for power quality estimation problems, where voltage signals often contain unmodeled disturbances and stochastic noise.

The H-$\infty$ estimator involves simple steps with low computational complexity, making it well-suited for practical implementation. At each time step, the estimated envelopes are fed into the ADALINE network to estimate the flicker components. This process is detailed in the following subsection.

\subsection{Flicker Components Estimation using ADALINE}

\begin{figure}[h]
\centering
\includegraphics[scale=0.6]{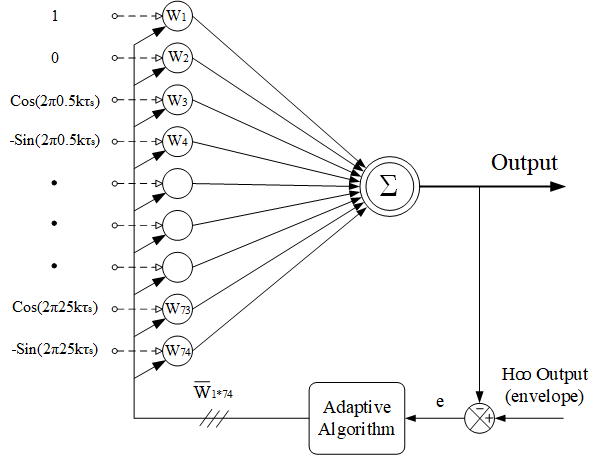}
\caption{Block diagram of the ADALINE neural network used for flicker frequency estimation. The input consists of sinusoidal basis functions covering a predefined flicker frequency range. Each input is weighted by a corresponding coefficient \( W_i \), and the weighted sum forms the ADALINE output. The output is compared against the reference envelope obtained from the H-$\infty$ filter. The resulting error \( e \) is used in the Least Mean Squares (LMS) adaptive algorithm to update the weights \( \vec{W} \) in real time, minimizing the squared error and allowing the network to track multiple flicker components.}
\label{fig1}
\end{figure}

The envelopes, as time-varying amplitudes, are extracted using Equation~\ref{eq12} at each time step. These envelopes are then fed into an online ADALINE neural network for spectral decomposition. The ADALINE network offers the advantage of a simple structure without requiring a learning phase, making it ideal for fast estimation processes. Moreover, the developed algorithm relies on a minimal number of tuning parameters, further simplifying its implementation. The schematic of the ADALINE network is illustrated in Fig.~\ref{fig1}. The target states for ADALINE are the flicker amplitudes at frequencies within the range \( f \pm F_i \) Hz, where \( f \) is the power frequency, and \( F_i \) is specified by IEC standard 61000-4-15, as detailed in Table~\ref{tab1}.

\begin{table}[h]
\centering
\caption{Flicker Frequencies \( F_i \) in Hz Based on IEC Standard 61000-4-15}
\label{tab1}
\begin{tabular}{c c c c c c c}
\hline
\multicolumn{6}{c}{Flicker Frequencies \( F_i \) (Hz)} \\ 
\hline
0.5 & 3.5 & 6.5 & 10 & 14 & 20 \\ \hline
1 & 4 & 7 & 10.5 & 15 & 21 \\ \hline
1.5 & 4.5 & 7.5 & 11 & 16 & 22 \\ \hline
2 & 5 & 8 & 11.5 & 17 & 23 \\ \hline
2.5 & 5.5 & 8.8 & 12 & 18 & 24 \\ \hline
3 & 6 & 9.5 & 13 & 19 & 25 \\ \hline
\end{tabular}
\end{table}

To estimate the flicker amplitudes, the extracted envelope from the H-$\infty$ filter is modeled, with the flicker frequencies selected from Table~\ref{tab1}. The network consists of 74 inputs representing DC values. 

The 74 frequencies in Table~\ref{tab1} are selected based on IEC Standard 61000-4-15, which defines the flicker frequency range most relevant to human visual perception (0.5–25 Hz). These form a fixed grid of basis functions used by the ADALINE network to estimate flicker amplitudes efficiently. Although the grid is discrete, the adaptive nature of the algorithm allows it to approximate off-grid components through spectral leakage into adjacent frequencies. This offers a practical balance between resolution and computational simplicity without requiring dynamic reconfiguration.

\begin{equation}
x_{ad}^k = [1, 0, \cos(2\pi 0.5 k \tau_s) -\sin(2\pi 0.5 k \tau_s) \dots \cos(2\pi 25 k \tau_s) -\sin(2\pi 25 k \tau_s)]^T
\label{eq13}
\end{equation}

The weight vector for the \( n \)-th envelope at time step \( k \) is given by:

\begin{equation}
w_{n}^k = [w_{1n}^k, w_{2n}^k, \dots, w_{73n}^k, w_{74n}^k]
\label{eq14}
\end{equation}

These weights are updated using the Widrow–Hoff rule.

\begin{equation}
    w_{n}^{k+1} = w_{n}^k + \frac{\theta^k e_{n}^k x_{ad}^{k^T}}{\lambda + x_{ad}^{k^T} x_{ad}^k}
    \label{eq15}
\end{equation}

where \( e_{n}^k \) is the tracking error and \( \lambda \) is a small quantity to avoid the denominator from being zero. The adaptive learning factor \( \theta^k \) is calculated as follows:

\begin{equation}
\theta^k = \frac{\theta^0}{1 + \frac{k}{\beta}}
\label{eq16}
\end{equation}

In Equation~\ref{eq16}, \( \beta \) is a constant that determines the decay rate of the initial learning rate \( \theta^0 \). In this form, the final weights encapsulate the flicker information. Finally, the harmonic and flicker amplitudes are extracted using the following formula:

\begin{equation}
V_n = \sqrt{w_{1n}^2 + w_{2n}^2} \quad \text{and} \quad \frac{\Delta V_i}{2V_t} = \frac{\sqrt{w_{(2i+1)n}^2 + w_{(2i+2)n}^2}}{V_n}, \quad i = 1, 2, \dots, F
\label{eq17}
\end{equation}

The final outputs of the ADALINE network are the relative flicker amplitudes 
$\Delta V_i / V_t$ corresponding to the standard flicker frequencies $F_i$. 
These represent the physical quantities of flicker defined in IEC 61000-4-15 
and are directly used to compute the instantaneous flicker sensation ($S$) 
and other standardized indices. Thus, the proposed method does not only 
track frequency content but explicitly estimates the relative amplitude of 
flicker components that have a direct physical and perceptual interpretation.

\section{Results}

A series of simulations were performed to assess the effectiveness of the proposed method in estimating flicker components. For this purpose, the waveform model depicted in Equation~\ref{eq2} was used. The waveform's frequency content aligns with the bus voltage of industrial loads, which typically includes a combination of the 1st to 11th harmonics. Additionally, significant flicker components based on the IEC Standard 61000-4-15 were incorporated into the waveform. To highlight the robustness of the proposed approach, the flicker estimation problem was applied to a highly distorted waveform containing five harmonic components and additive high-frequency noise.

Appropriate parameter values for the proposed method were chosen. The selected parameters for the H-$\infty$ filter and the ADALINE neural network are presented in Table~\ref{tab2}.

\begin{table}[H]
\small
\centering
\caption{Parameters for H-$\infty$ filter - ADALINE neural network simulations}
\label{tab2}
\begin{tabular}{|c c|c c|c c|}
\hline
\multicolumn{2}{|c|}{\textbf{H$_\infty$ method}} & \multicolumn{2}{c|}{\textbf{ADALINE method}} & \multicolumn{2}{c|}{\textbf{Simulation}} \\ \hline
\( P^0 \)  & \( 10^3 \cdot I_{2N \times 2N}^* \) & \( w^0 \) & \( 1_{2i+2}^{**} \) & \( \tau_s \) & \( 1/1200 \) \\ \hline
\( \hat{x}_0 \) & \( 1_{2N}^{**} \) & \( \theta_0 \) & \( 5 \) & \( \sigma \) & \( 0.02 \) \\ \hline
\( \alpha \) & \( 8 \) & \( \lambda \) & \( 0.0001 \) &  &  \\ \hline
\( f \) & \( 50 \) & \( \beta \) & \( 1000 \) &  &  \\ \hline
\( R \) & \( 0.007 \) &  &  &  &  \\ \hline

\end{tabular}
\end{table}

\( I^* \) represents the identity matrix, and \( 1^{**} \) is a vector of ones. The selected parameters remain constant throughout all simulations. The key software simulations considered in this study are as follows:

\begin{enumerate}
    \item Simulated waveform with one power harmonic and one flicker component.
    \item Distorted waveform with 5 power harmonics and 36 flicker components. This test examines the proposed method's capability to track flicker in the presence of multiple harmonics.
    \item Monte Carlo (MC) simulation for analyzing the proposed method. This test demonstrates the method's generalization ability, as harmonics and flicker amplitudes are randomly selected in the signal model.
\end{enumerate}

\subsection{Estimation of a Waveform with Power Frequency and One Flicker Component}

As a representative case study, a waveform containing a power frequency and a single flicker component is simulated. The power frequency is 50 Hz, and it is estimated using a PLL from the MATLAB toolbox. To evaluate the performance of the proposed method, Gaussian noise is added to the signal as follows:

\begin{equation}
Z^k = 1.5 \cos(2\pi 50 k \tau_s + 80^\circ) \times \left\{ 1 + \frac{\Delta V_i}{2V_t} \cos(2\pi F_i k \tau_s + \theta_i) \right\} + 0.02 \cdot \text{randn}^k
\label{eq18}
\end{equation}

\begin{figure}[h]
    \centering
    \includegraphics[width=0.7\linewidth]{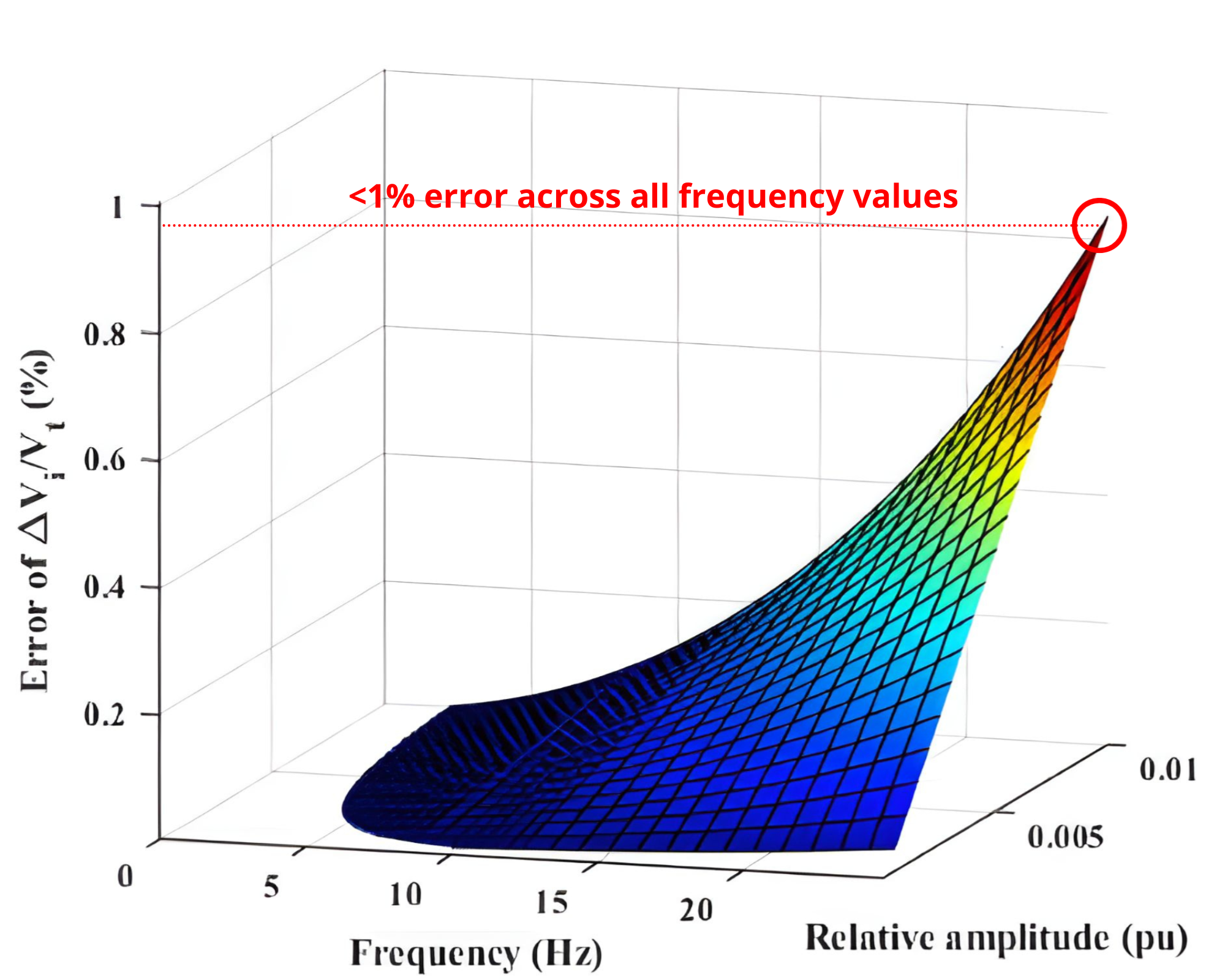}
    \caption{Absolute error of \((\Delta V_i)/V_t\)  estimation, waveform with power frequency and one flicker component.}
    \label{fig2}
\end{figure}

\begin{figure}[h]
    \centering
    \includegraphics[width=0.7\linewidth]{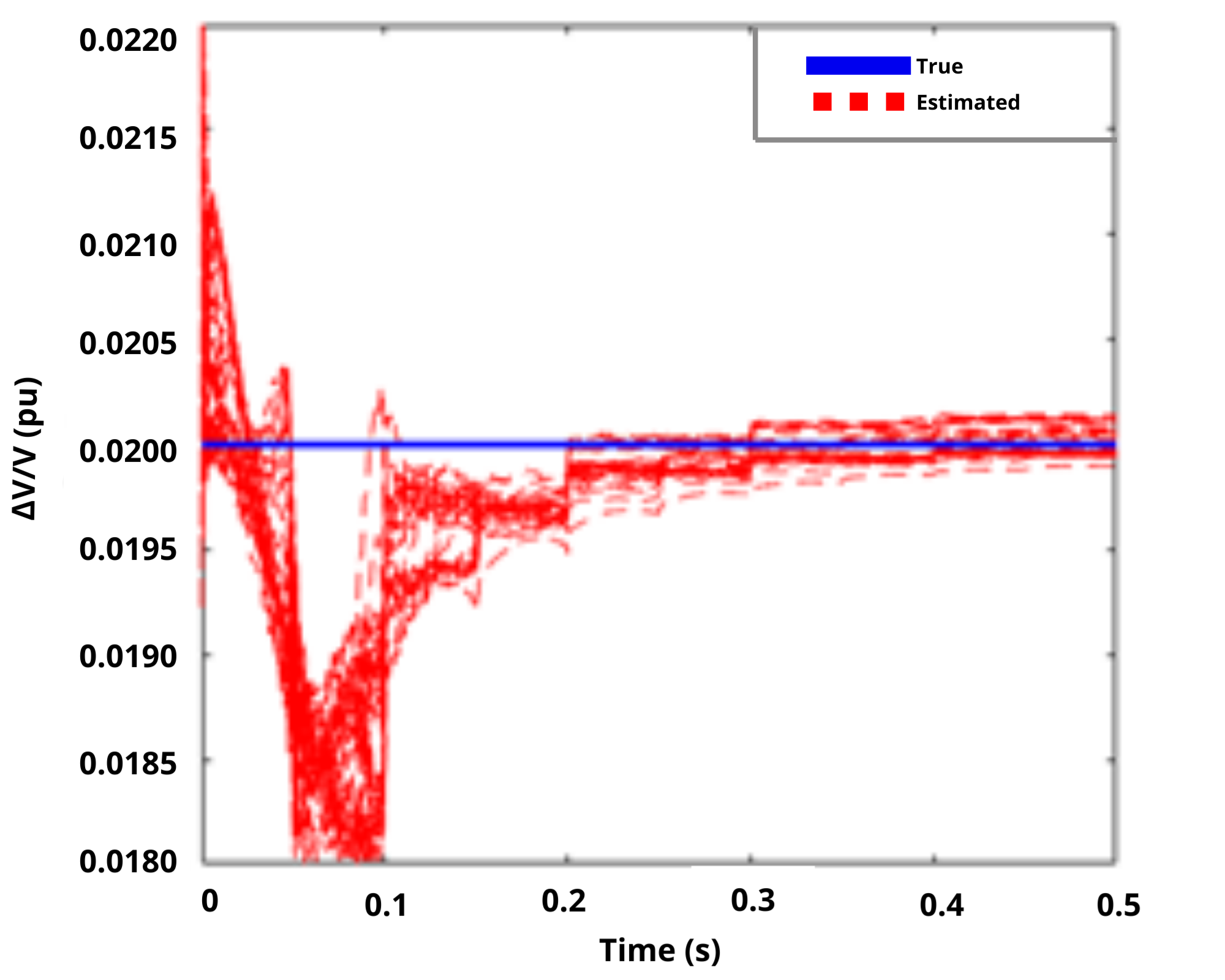}
    \caption{Trend of convergence of the proposed method. The blue line represents the true relative flicker amplitude \(\Delta V/V = 0.02\) pu. The red lines show the estimated flicker amplitude computed by the ADALINE neural network over time, across multiple Monte Carlo runs. The plot illustrates the convergence behavior of the estimator under a waveform and one flicker component.}

    \label{fig3}
\end{figure}

Several simulations are conducted, with one of the flicker frequencies from Table~\ref{tab1} injected into the waveform. The relative voltage amplitude of the flicker component, \( \frac{\Delta V_i}{V_t} \), is randomly selected within the range [0.001, 0.02]. Additionally, \( \theta_i \) values are randomly generated within the [0, $90^\circ$] boundary. The proposed method estimates the relative amplitude of the flicker components. The final estimated parameters are compared with the corresponding true values, and the absolute error is calculated. The results of the error analysis for different simulations are presented in Fig.~\ref{fig2}. Although the absolute error of the estimation increases with amplitude and frequency, the algorithm consistently estimates the amplitudes with less than 1\% error across all frequency values. The convergence trend is also examined, as shown in Fig.~\ref{fig3}, where acceptable results are achieved for all flicker frequencies within 0.3 seconds.

\subsection{Estimation of distorted waveform with 5 power harmonics and 36 flicker components}

Due to non-sinusoidal current anomalies present across the power lines, the actual bus voltages exhibit a distorted nature, which can be represented as a waveform comprised of a series of harmonics. In this section, the flicker estimation problem is addressed by employing a modulated waveform containing multiple power harmonics. Additionally, to assess the performance of the proposed method under the influence of numerous flicker frequencies, the fluctuating voltage waveform encompasses all frequencies defined in the IEC 61000-4-15 standard. Thus, the fluctuating signal in Equation~\ref{eq2} is formulated as follows \cite{b15, b19, b39, b40}:

\begin{equation}
\begin{split}
Z^k = \left[ 1.5 \cos\left( 2\pi 50 k \tau_s + 80^\circ \right) 
+ 0.5 \cos\left( 2\pi 150 k \tau_s + 60^\circ \right) \right. \\
+ 0.2 \cos\left( 2\pi 250 k \tau_s + 45^\circ \right) \\
+ 0.15 \cos\left( 2\pi 350 k \tau_s + 36^\circ \right)
\left. + 0.1 \cos\left( 2\pi 550 k \tau_s + 30^\circ \right) \right] \\
\left\{ 1 + \sum_{i=1}^{F} \frac{\Delta V_i}{2 V_t} 
\cos\left( 2\pi F_i k \tau_s + \theta_i \right) \right\} 
+ 0.02 \cdot \text{randn}^k    
\end{split}
\label{eq19}
\end{equation}

\begin{figure}[h]
    \centering
    \includegraphics[scale=0.4]{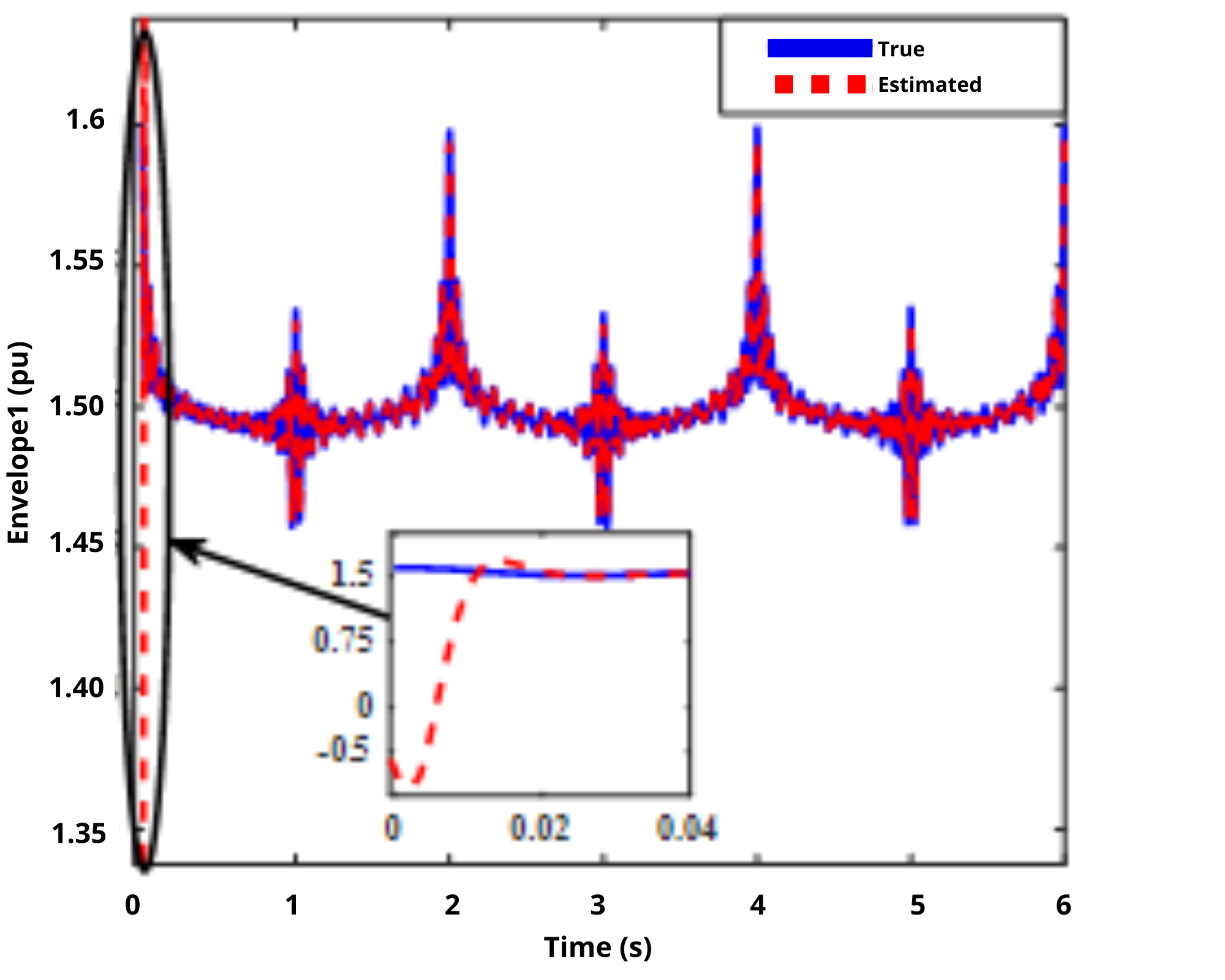}
    \caption{Estimation of envelope 1 using H-infinity filter.}
    \label{fig4}
\end{figure}

\begin{figure}[h]
    \centering
    \includegraphics[scale=0.4]{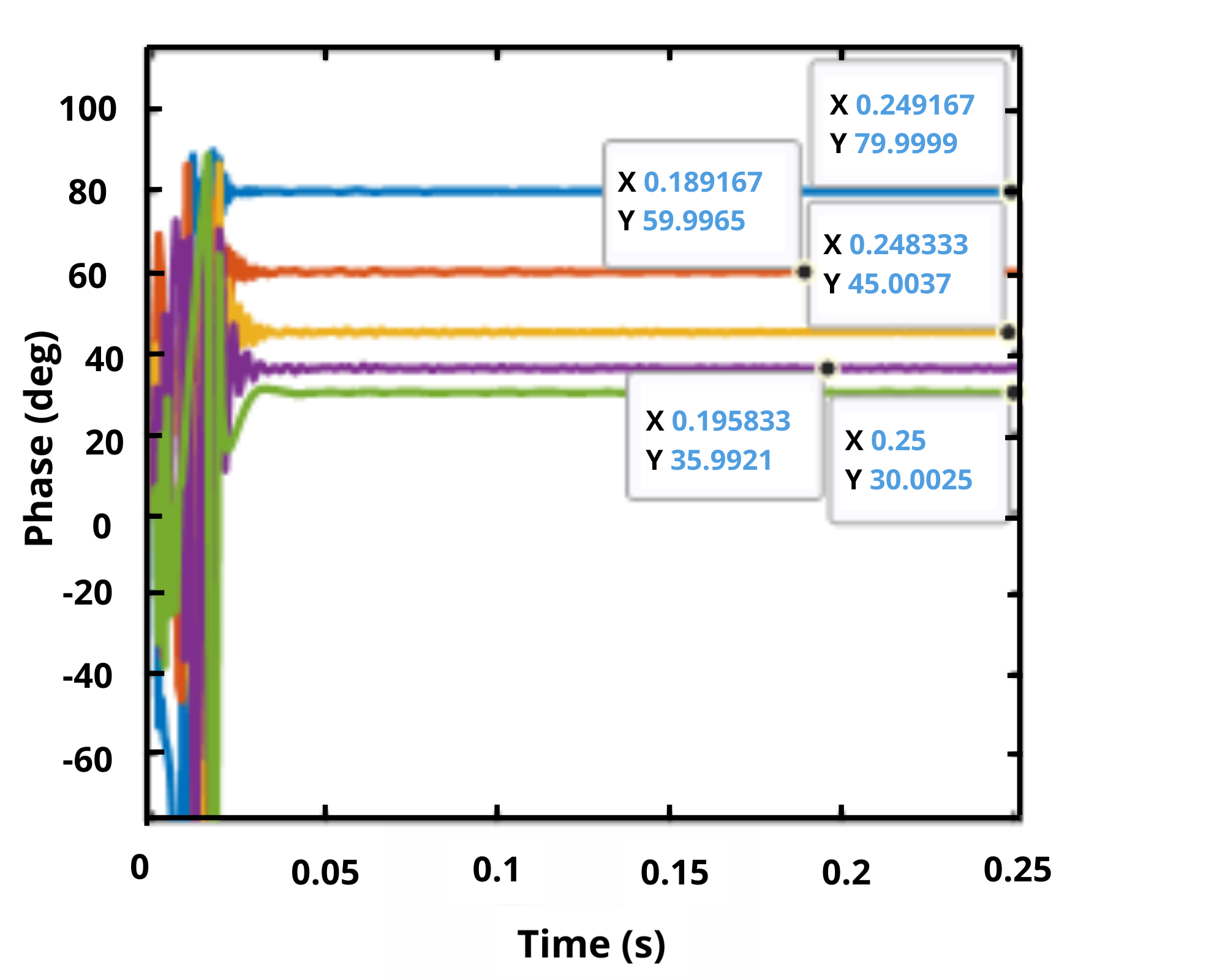}
    \caption{Estimation of power harmonic phases using H-infinity filter. In \textcolor{blue}{blue the main components}, in \textcolor{red}{red the 3rd component}, in \textcolor{orange}{orange the 5th component},in \textcolor{purple}{purple the 7th component} and in \textcolor{green}{green the 11th component.}}
    \label{fig5}
\end{figure}

The flicker frequencies \( f_{i} \) and their corresponding amplitudes \( \frac{\Delta V_i}{V_t} \) are presented in Table~\ref{tab3}. The chosen amplitudes ensure a unity instantaneous flicker sensation in 230V/50Hz systems \cite{b36}. At each moment \( k \), the H$_\infty$ filter extracts the signal envelopes. In the case under study, 5 harmonics are introduced into the simulation signal, necessitating the estimation of five envelopes by the H$_\infty$ filter. The estimation results for envelope 1 are illustrated in Fig.~\ref{fig4}, where the proposed algorithm converges to the true value in less than one power cycle (0.02s). The remaining estimated envelopes display a similar tracking behavior. According to Equation~\ref{eq12}, the phases of the power harmonic components are also determined by the H$_\infty$ filter. The results of the phase estimation for the power harmonics are shown in Fig.~\ref{fig5}. Due to the PLL error in detecting the exact frequency, the convergence of the harmonic phases to the reference value is slightly slower compared to that of the envelopes. Nonetheless, all phases converge within 0.03s.Despite these challenges, the H-$\infty$ filtering approach maintains stable estimates due to its inherent robustness against measurement and modeling uncertainties. Nevertheless, careful tuning of PLL parameters or inclusion of advanced PLL structures could further minimize such effects.

\begin{figure}[h]
    \centering
    \includegraphics[scale=1]{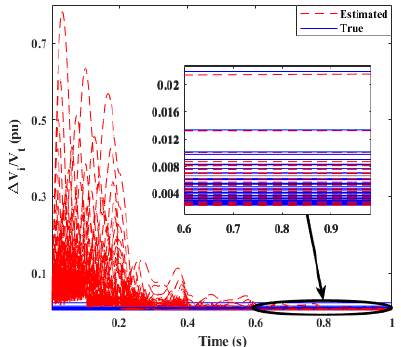}
    \caption{Estimation of flicker amplitudes $\frac{\Delta V_i}{V_t}$ using ADALINE.}
    \label{fig6}
\end{figure}

\begin{figure}[h]
    \centering
    \includegraphics[scale=1]{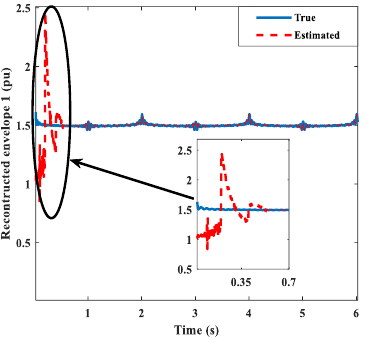}
    \caption{Envelope 1 reconstruction using estimated flicker amplitudes.}
    \label{fig7}
\end{figure}

In the second stage at each time step \( k \), the envelopes estimated by the H$_\infty$ filter are used as inputs to the ADALINE neural network to estimate the flicker components \( \frac{\Delta V_i}{V_t} \). Fig.~\ref{fig6} shows the tracking behavior of flicker amplitudes using the ADALINE network. Given that the initial weights of the network are selected randomly, the estimation results exhibit initial overshoots. The proposed hybrid algorithm estimates harmonic and flicker components through separate processes, allowing the extraction of parameters with varying scales independently. This approach reduces the number of iterations required for convergence, with the data processing of the distorted signal completed within 0.6s. Using the estimated amplitudes, the envelopes are reconstructed. Specifically, the reconstructed envelopes closely match their actual values once the estimated flicker components converge to their true values. Fig.~\ref{fig7} illustrates the quality of the envelope 1 reconstruction based on the ADALINE estimator results, and this trend can be generalized to the other envelopes.

 \begin{table}[h]
\centering
\caption{Flicker frequencies \( F_i \) and corresponding relative amplitudes for distorted waveform estimation}
\label{tab3}
\begin{tabular}{cccccc}
\hline
\(F\) & \(\frac{\Delta V}{V_t}\) & \(F\) & \(\frac{\Delta V}{V_t}\) & \(F\) & \(\frac{\Delta V}{V_t}\) \\
\hline
0.5  & 0.0234  & 6.5  & 0.003    & 14   & 0.00388 \\
1    & 0.01432 & 7    & 0.0028   & 15   & 0.00432 \\
1.5  & 0.0108  & 7.5  & 0.00266  & 16   & 0.0048  \\
2    & 0.00882 & 8    & 0.00256  & 17   & 0.0053  \\
2.5  & 0.00754 & 8.8  & 0.0025   & 18   & 0.00584 \\
3    & 0.00654 & 9.5  & 0.00254  & 19   & 0.0064  \\
3.5  & 0.00568 & 10   & 0.0026   & 20   & 0.007   \\
4    & 0.005   & 10.5 & 0.0027   & 21   & 0.0076  \\
4.5  & 0.00446 & 11   & 0.00282  & 22   & 0.00824 \\
5    & 0.00398 & 11.5 & 0.00296  & 23   & 0.0089  \\
5.5  & 0.0036  & 12   & 0.00312  & 24   & 0.00962 \\
6    & 0.00328 & 13   & 0.00348  & 25   & 0.01042 \\
\hline
\end{tabular}
\end{table}

Table \ref{tab4} summarizes the estimation errors of the first envelope due to phase-locked loop (PLL) inaccuracies, under conditions with and without fundamental frequency deviations.

\begin{table}[h]
\small
    \centering
    \caption{Error of envelope 1 estimation due to PLL errors}
    \label{tab4}
    \begin{tabular}{c c c c}
    \hline
    \multicolumn{2}{c}{\textbf{Without frequency deviation}} & \multicolumn{2}{p{6cm}}{\textbf{With frequency deviation of $\pm$ 0.5 Hz}} \\ \hline
    \textbf{Mean} & \textbf{Variance} & \textbf{Mean} & \textbf{Variance}  \\ \hline
    \textbf{\(5.352e^{-6}\)} & $4.183e^{-6}$ & $7.266e^{-5}$ & $8.419e^{-5}$ \\ \hline
    \end{tabular}
\end{table}

As shown in Table \ref{tab4}, the estimation error is negligible in the absence of frequency deviations, with a mean error of $5.352\times10^{-6}$ and variance of $4.183\times10^{-6}$. However, introducing a ±0.5 Hz frequency deviation significantly increases both the mean and variance of the error, by more than an order of magnitude. This highlights the sensitivity of PLL-based approaches to frequency variations, reinforcing the advantage of the proposed H-$\infty$ filter framework, which remains robust under similar conditions.

\subsection{Impact of PLL Errors and Optimization Strategies}

As shown in Table~\ref{tab4}, PLL inaccuracies can introduce estimation errors, particularly under frequency deviation conditions. While the proposed H-$\infty$–ADALINE framework demonstrates robustness, its accuracy ultimately depends on the ability of the PLL to track the fundamental frequency. To further mitigate this source of error, adaptive PLL strategies may be employed. For instance, second-order generalized integrator PLLs (SOGI-PLLs) and adaptive notch filter PLLs offer faster convergence and enhanced tracking under dynamic frequency deviations \cite{pll}. 

Additionally, the tradeoff between PLL bandwidth and noise sensitivity must be carefully tuned to balance tracking precision and robustness. Integrating such advanced PLL structures with the proposed method is expected to further improve reliability under non-stationary conditions, and will be explored in future work.

\subsection{Monte Carlo (MC) analysis and results discussion}

\begin{figure}[h]
    \centering
    \includegraphics[scale=0.3]{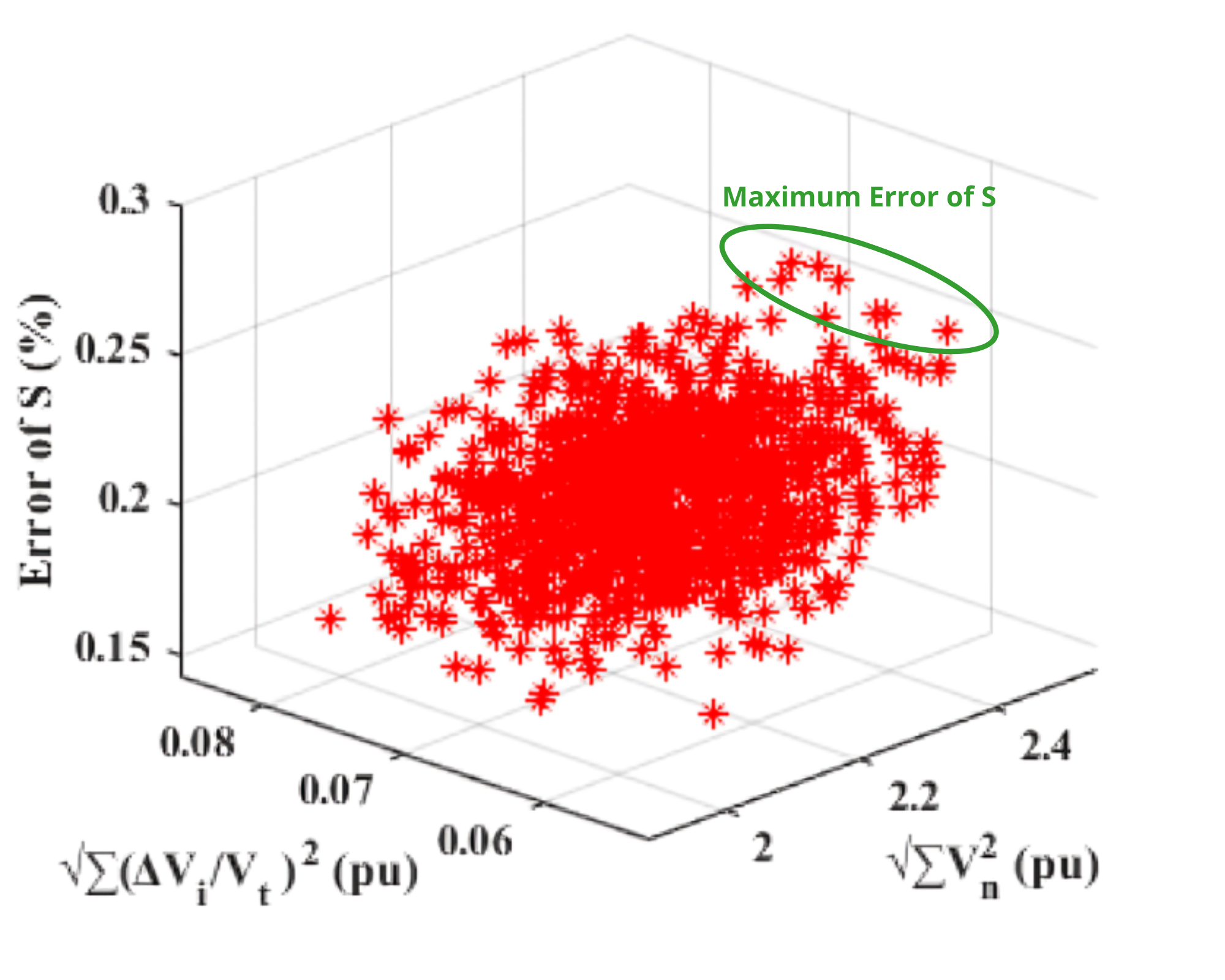}
    \caption{Results of MC analysis for S level estimation. }
    \label{fig8}
\end{figure}

MC-based analyses encompass a variety of problem-solving techniques that utilize random numbers as inputs to generate output statistics \cite{b15, b33, b41,b42, b43, b44}. One of the primary applications of MC methods is to validate the performance of estimation algorithms in the presence of uncertainties. In this study, the MC method consists of a series of simulations, where each simulation employs random values for harmonics and flicker amplitudes in the waveform defined by Equation~\ref{eq2}. These random values are uniformly sampled from the intervals [0.8, 1.2] and [0, 0.02] for harmonic and flicker amplitudes, respectively. The proposed estimation approach is then applied to estimate the flicker amplitudes. The instantaneous flicker sensation (S), which serves as the evaluation index for flicker, is computed using the estimated flicker amplitudes to assess the estimation quality. The contribution of each flicker frequency to \( S \) can be expressed as follows:

\begin{equation}
S_i = \frac{\left(\frac{\Delta V_i}{V_t}\right)}{\left(\frac{\Delta V_i}{V_t}\right)_{\text{IEC}}}
\label{eq20}
\end{equation}

where \( \left(\frac{\Delta V_i}{V_t}\right)_{\text{IEC}} \) is derived from Table~\ref{tab3} for the flicker frequency \( F_i \). The total estimated sensation \( S \) is obtained by summing the contributions of the individual flicker frequencies \( S_i \). Finally, the estimated sensation is compared with the true sensation, and the error percentage is calculated as the output of the MC simulation. Fig.~\ref{fig8} depicts the distribution of the error in \( S \) with respect to the root of the sum of squared harmonic amplitudes \( \sqrt{\sum_{n=1}^{N} V_n^2} \) and the root of the sum of squared flicker amplitudes \( \sqrt{\sum_{i=1}^{F} \left(\frac{\Delta V_i}{V_t}\right)^2} \).

\begin{figure}[h]
    \centering
    \includegraphics[scale=0.4]{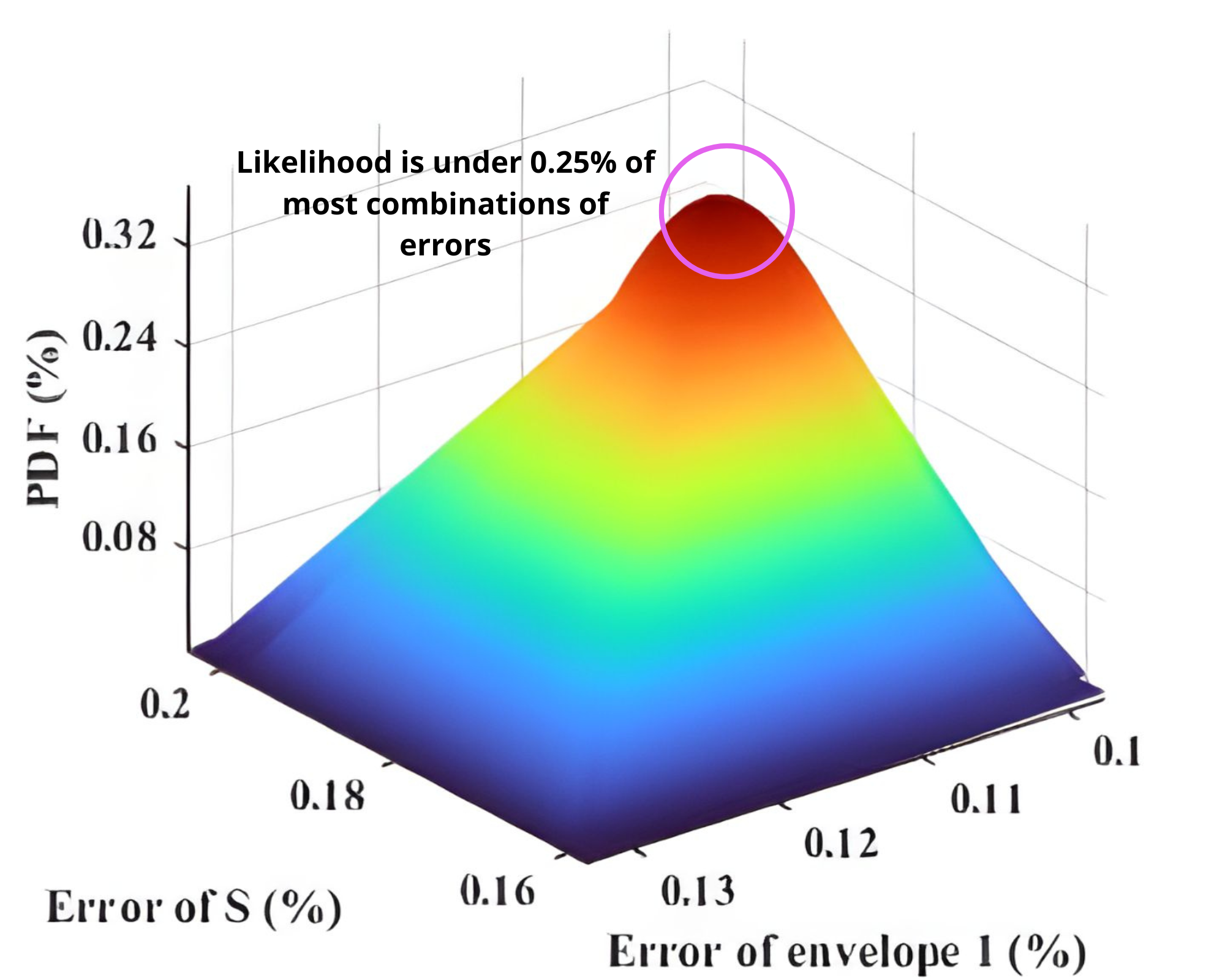}
    \caption{MC statistical analysis for frequency and time domains. }
    \label{fig9}
\end{figure}

As shown in Fig.~\ref{fig8}, the error in \( S \) remains below 0.25\% for all random combinations of harmonic and flicker amplitudes. Additionally, the convergence behavior of the algorithm in both the frequency and time domains is evaluated through multiple MC runs. In this context, a 2D statistical distribution of the error in \( S \) and the error in envelope 1 is examined. The results, illustrated in Fig.~\ref{fig9} and summarized in Table~\ref{tab5}, not only highlight the accuracy of the proposed algorithm in extracting the frequency-domain characteristics of the studied waveform (i.e., \( S \) values) but also demonstrate its precision in estimating time-domain features (such as the envelope values).

\begin{table}[H]
\small
\centering
\caption{ NUMERICAL ANALYSIS OF MC OUTPUT \\
*Standard deviation}
\label{table:error_analysis}
\begin{tabular}{lcc}
\hline
 & \textbf{Error of S (\%)}  & \textbf{Error of envelope 1 (\%)} \\
\hline
\textbf{Mean} & 0.1832 & 0.1114 \\
\textbf{STD*} & 0.0021 & 0.0038 \\
\hline
\end{tabular}
\label{tab5}
\end{table}

\section{Discussion}

\subsection{Theoretical and practical insights}

\begin{figure}[H]
    \centering
    
    \begin{subfigure}[b]{0.5\textwidth}
        \centering
        \includegraphics[width=\textwidth]{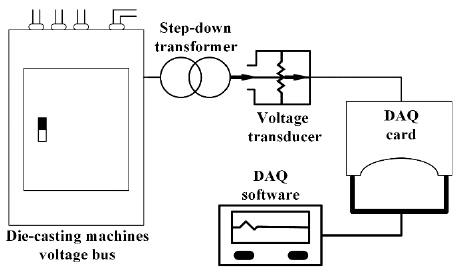}  
        \caption{Testbench schematic}
        \label{fig10a}
    \end{subfigure}
    \hfill  
    
    \begin{subfigure}[b]{0.5\textwidth}
        \centering
        \includegraphics[width=\textwidth]{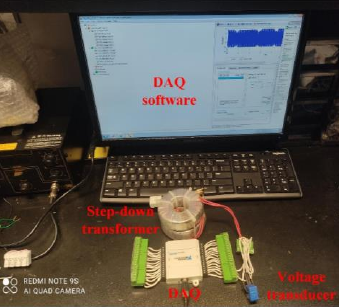}  
        \caption{Real-world components}
        \label{fig10b}
    \end{subfigure}

    \caption{Data logger setup dedicated to the experimental test.}
    \label{fig10}
\end{figure}

\begin{figure}[H]
    \centering
    \includegraphics[width=0.7\linewidth]{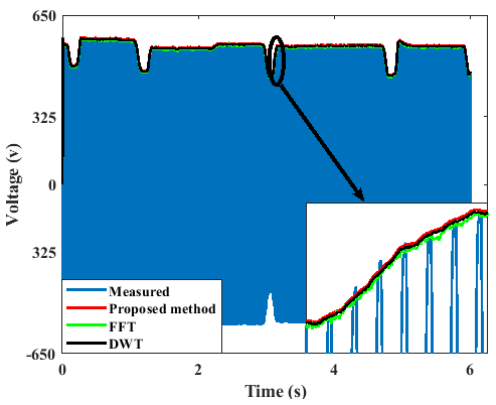}
    \caption{Experimental signal analysis using different methods for validation.}
    \label{fig11}
\end{figure}

The theoretical performance of the proposed method has been previously validated through various simulation tests. In this section, the algorithm's applicability in real-world scenarios is examined. Additionally, conventional Fast Fourier Transform (FFT) and state-of-the-art Discrete Wavelet Transform (DWT) methods, provided by MATLAB toolboxes, are utilized for flicker component estimation under the same experimental conditions, and their results are compared. A hardware setup was designed to log real data from a fluctuating voltage source. The voltage waveform is derived from a 400V infinite bus voltage powering die-casting machines with operational forces ranging from 160 to 1000 tons, electrostatic painting machines, and oil pumps at Mazinoor Lighting Industries Inc. Due to the sudden load changes of the die-casting machines, the input electrical current experiences abrupt variations, leading to fluctuations in the supply voltage. Die-casting machines were selected due to their frequent and significant transient load changes, effectively illustrating the robustness and real-time capability of the proposed method. To demonstrate broader applicability, further testing on inverter-based loads, such as variable-frequency drives (VFDs), or induction motors—common sources of harmonics and flicker in industrial environments—is recommended in future studies. The machine's input voltage is measured using a fast-response voltage transducer (LV 25-P), and the data is logged via an analog-to-digital NI USB-6009 DAQ card at a sampling rate of 1200 Hz for a duration of 6 seconds. The experimental setup is shown in Fig.~\ref{fig10a}, while the data logging configuration is presented in Fig.~\ref{fig10b}. The data is then processed offline using the proposed algorithm, as well as the FFT and DWT methods, within MATLAB. The computations were performed on a computer equipped with a Core i5, 3.1 GHz processor, and 8GB of RAM. Fig.~\ref{fig11} presents the graphical results for envelope 1 across the three methods. Due to abrupt load changes, the measured signal contains significant notches, but the estimated envelope 1 closely follows the measured signal's envelope. For more detailed analysis, the undervoltage section of the test waveform is magnified, highlighting the superior performance of the proposed algorithm and DWT method over FFT in both steady-state and transient conditions. The estimated signal, reconstructed by applying the estimated flicker components and harmonic parameters, is compared with the measured signal. The error statistics for the final two seconds (4s to 6s) of data for the FFT, DWT, and proposed methods are presented in Table~\ref{tab6}. Despite the favorable graphical results, the numerical indices reveal that the DWT method exhibits lower accuracy in practical situations. The primary reason for the reduced accuracy of the DWT time-frequency domain method lies in the presence of high-frequency noise elements in real data, which cause aliasing effects.

For the second test, the stable feeder was disconnected, and the load was powered by a diesel emergency generator. Real-time measurements indicate a frequency deviation of ±0.25Hz during die-casting machine operation. In such conditions, the PLL results for power frequency estimation are biased due to interference (noise or narrowband interference), affecting the proposed method's accuracy. While all methods demonstrate reduced accuracy under these conditions, the proposed method remains stable by decomposing the voltage fluctuations and harmonic components, thereby mitigating frequency interference effects. As a result, the impact of PLL errors is reduced in the outputs. However, the combination of aliasing and the picket-fence effect degrades the performance of the FFT and DWT methods. The numerical results are shown in Table~\ref{tab6}, which also includes an analysis of the computational burden of the three methods during the final two seconds of processed data. The proposed method demonstrates the best computational performance under normal conditions (stable feeder). Nevertheless, in the presence of frequency deviation, the required processing time is slightly higher than that of the FFT method, as additional iterations are needed to achieve acceptable results. Despite the accurate results, the DWT method incurs a significant computational burden, which limits its feasibility for real-time applications. The parameters in Table~\ref{tab6} underscore the proposed hybrid method's ability to provide both accurate and efficient flicker component estimation in practical situations.

While this study focuses primarily on flicker estimation, the hybrid method’s underlying adaptive structure—combining H-$\infty$ filtering and ADALINE networks—could potentially be adapted to estimate other power disturbances such as voltage sags, swells, and interharmonics. Future research should verify this generalization explicitly, potentially adjusting the model formulation to account for additional frequency and amplitude variations characteristic of these disturbances.

\begin{table}[H]
    \small
    \centering
    \caption{Numerical analysis of practical data}
    \begin{tabular}{c p{2cm} c c p{2cm}}
    \hline
    \textbf{Method} & \textbf{Feature} & \textbf{MSE} & \textbf{Variance} & \textbf{Computational Time} \\ \hline
    \multirow{3}{*}{Infinite Feeder} & \textbf{Proposed Method}  & $7.776e^{-6}$ & $5.911e^{-6}$ & 0.3812s \\ \cline{2-5}
     & FFT&$ 4.367e^{-5}$ & $1.019e^{-4}$ & 0.4182s \\ \cline{2-5}
     & DWT& $7.8115e^{-6}$ & $7.381e^{-5} $& 0.8338s \\ \hline
    \multirow{3}{*}{Emergency Diesel} & \textbf{Proposed Method}  & $1.912e^{-4}$ & $4.54e^{-4}$ & 0.4290s \\ \cline{2-5}
     &FFT & $3.007e^{-3}$ & $5.672e^{-4}$ & 0.4194s \\ \cline{2-5}
     &DWT & $2.716e^{-4}$ & $4.705e^{-4}$ & 0.8627s \\ \hline
    \end{tabular}
    \label{tab6}
\end{table}

\subsection{Parameter Selection and Optimization}

The performance of the proposed H-$\infty$–ADALINE method depends on several key parameters. 
For the H-$\infty$ filter, the weighting matrices and regularization parameter $\lambda$ were selected to balance robustness against measurement noise with stability guarantees, following the standard design procedure. The chosen values ensure bounded estimation error while maintaining responsiveness under dynamic disturbances. 

For the ADALINE network, the initial learning factor $\theta^0$ and decay constant $\beta$ (Equation~\ref{eq16}) were tuned through sensitivity analysis. Larger values of $\theta^0$ improve convergence speed but risk instability, while smaller values enhance stability at the cost of slower adaptation. In our study, the selected parameters provided a practical tradeoff, validated through Monte Carlo simulations based on IEC 61000-4-15 test cases. 

Although manual tuning yielded satisfactory results, systematic optimization methods such as Bayesian optimization, genetic algorithms, or adaptive gain scheduling could further enhance parameter robustness, especially for large-scale or highly dynamic systems. Incorporating such automated tuning methods will be considered in future work.

\subsection{Scalability and Feasibility Considerations}

While the proposed hybrid H-$\infty$–ADALINE approach has been validated within the IEC-defined flicker frequency range (0.5–25 Hz), its computationally efficient recursive structure suggests promising scalability to larger-scale distribution systems. The H-$\infty$ filter requires only matrix updates at each iteration, and the ADALINE network relies on a single-layer architecture without offline training, both of which are compatible with real-time deployment in power quality monitoring devices.

Nevertheless, large interconnected power systems introduce additional challenges such as complex harmonic interactions, interharmonics, and asynchronous disturbances that may affect estimation accuracy. Moreover, extending the analysis beyond the 25 Hz range (e.g., into the supraharmonic domain) may require modifications to the ADALINE basis functions or the inclusion of multi-resolution processing to capture higher-frequency dynamics. 

Therefore, while the proposed method is theoretically well-suited for extension to large-scale systems and higher frequency bands, future work will focus on empirical validation through field data and adaptation of the algorithm for enhanced robustness in complex grid scenarios.

\section{Conclusion}

This paper introduces a novel algorithm for the estimation of flicker frequency components in voltage waveforms, addressing the inherent complexity arising from the multitude of frequencies involved. The proposed method effectively decomposes the voltage waveform into envelope and harmonic components, significantly simplifying the estimation process. Envelopes are accurately extracted using a robust $H-\infty$ filter, after which an online ADALINE neural network efficiently estimates the flicker content within these envelopes. Additionally, the study presents a new voltage waveform model that incorporates flicker, harmonics, and white noise to comprehensively assess algorithm performance.

Graphical and numerical results demonstrate the effectiveness of the proposed method in accurately tracking power system flickers. Monte Carlo simulations further confirm its robustness against various harmonic and flicker scenarios aligned with IEC Standard 61000-4-15. Practical applicability has been validated using real-world fluctuating voltage data, showing that the proposed method outperforms conventional FFT and DWT methods in terms of Mean Square Error (MSE), variance, and computational efficiency.

The key strengths of this method include robustness against measurement noise, rapid convergence suitable for real-time implementations, and computational efficiency ideal for hardware-constrained environments. Collectively, these attributes render the proposed method highly suitable for industrial power quality monitoring.

Despite its advantages, the method exhibits certain limitations. The fixed frequency grid, although compliant with standards, may constrain accuracy for off-grid frequencies. Additionally, the validation primarily focuses on die-casting machine loads, which limits the generalizability to other industrial scenarios. The absence of modeling for modern lighting technologies, particularly LEDs, also represents a gap.

Future work could explore the impact of higher-order harmonics, flickers, and interharmonics. Testing other robust online neural networks, such as Online Recurrent Neural Networks (RNNs) or Echo State Networks, could further enhance prediction accuracy and computational speed. Additionally, extending the model to accurately capture flicker characteristics of modern LED lighting systems represents an important avenue for continued research.







\section*{Acknowledgement}

\section*{Author contributions: CRediT}

\end{document}